\documentstyle[11pt,epsf]{article}

\newcommand{\be}{\begin{equation}}
\newcommand{\ee}{\end{equation}}
\newcommand{\nn}{\nonumber}
\newcommand{\beba}{\begin{equation}\begin{array}{lcl}}
\newcommand{\eaee}{\end{array}\end{equation}}
\newcommand{\bea}{\begin{eqnarray}}
\newcommand{\eea}{\end{eqnarray}}
\newcommand{\ba}{\begin{array}}
\newcommand{\ea}{\end{array}}

\newcommand{\ns}{\normalsize}
\newcommand{\refs}[1]{(\ref{#1})}
\def\bal{{\mbox{\boldmath $\alpha$}}}

\def\bq{{\bf q}}
\def\bd{{\bf d}}

\def\a{\alpha}
\def\b{\beta}
\def\g{\gamma}

\def\d{\delta}
\def\e{\epsilon}

\def\f{\phi}

\def\l{\lambda}
\def\m{\mu}
\def\n{\nu}

\def\t{\tau}

\def\L{\Lambda}
\def\O{\Omega}

\begin{document}

%%%%%%%%%%%%%%%%%%%%%%%%%%%%%%%%%%%%%%%%%%%%%%%%%%%%%%%%%%%%%%%%%%%%%%%%%%%

\begin{titlepage}
\title{\hfill{\ns UPR-726T, IASSNS-HEP-96/118, PUPT-1667\\}
       \hfill{\ns hep-th/9611204\\[.1cm]}
       \hfill{\ns November 1996}\\[.8cm]
       {\Large\bf Stabilizing Dilaton and Moduli Vacua in String
                  and M--Theory Cosmology}}
\author{Andr\'e
        Lukas$^1$\setcounter{footnote}{0}\thanks{Supported by Deutsche
        Forschungsgemeinschaft (DFG) and
        Nato Collaborative Research Grant CRG.~940784.}~~,
        Burt A.~Ovrut$^1\; ^3$ and Daniel Waldram$^2$\\[0.5cm]
        {\ns $^1$Department of Physics, University of Pennsylvania} \\
        {\ns Philadelphia, PA 19104--6396, USA}\\[0.3cm]
        {\ns $^2$Department of Physics}\\
        {\ns Joseph Henry Laboratories, Princeton University}\\
        {\ns Princeton, NJ 08544, USA}\\[0.3cm]
        {\ns $^3$School of Natural Sciences, Institute for Advanced Study}\\
        {\ns Olden Lane, Princeton, NJ 08540, USA}}
\date{}
\maketitle

\begin{abstract} 

We show how non-trivial form fields can induce an effective potential
for the dilaton and metric moduli in  compactifications of type II
string theory and M--theory. For particular configurations, the
potential can have a stable minimum. In  cosmological
compactifications of type II theories, we demonstrate that, if the
metric moduli become fixed, this mechanism can then lead to the
stabilization of the dilaton vacuum. Furthermore, we  show that for
certain cosmological M--theory solutions, non-trivial forms lead to
the stabilization of moduli. We present a number of examples,
including cosmological solutions with two solitonic forms and examples
corresponding to the infinite throat of certain $p$--branes. 

\end{abstract}

\thispagestyle{empty}
\end{titlepage}

%%%%%%%%%%%%%%%%%%

\section{Introduction}

Recent advances in understanding the structure underlying string
theory have renewed interest in type II and eleven-dimensional
supergravities and the role played by the form-field degrees of
freedom in these theories. It is now understood that there are
strong-weak coupling dualities~\cite{dual} relating each of the known
supersymmetric string theories together with a putative theory in
eleven-dimensions, M--theory. This suggests that the corresponding
low-energy effective theories, which are supergravities, may be
directly relevant to particle physics and cosmology. A second
development has been the discovery of D--brane states in open string
theory as sources of Ramond-Ramond (RR) form field charge in type II
supergravities. These states have proved central to the recent
understanding of some of the statistical dynamics of black
holes~\cite{bhs}. 

Given this new perspective, it becomes interesting to ask what role
these form fields might play in compactifications of string theory and
M-theory. 
Phenomenologically, probably the most relevant
case in M--theory is the compactification on a Calabi-Yau three-fold cross the
orbifold $S_1/Z_2$, considered by Witten~\cite{witten}, which
describes the strong coupling limit of heterotic string
theory~\cite{hor_witt}. Some of the particle physics and cosmological
implications of this limit have been discussed by Banks and
Dine~\cite{banks_dine} and Horava~\cite{hor}. In 
each of these compactifications, the
three-form potential in eleven-dimensional supergravity is excited. In
fact, it can be shown that it cannot be set to
zero~\cite{witten}. Compactifications of type II theory with non-trivial
form fields have been considered in~\cite{tII}. 
In two recent papers~\cite{letter,paper}, we studied cosmological
solutions with compact, but dynamic, internal spaces and non-trivial form
fields. Assuming that the internal space was a product of maximally
symmetric subspaces, the general solutions are found to be closely
related to non-extremal black $p$-brane solutions~\cite{blackp},
though with the role of the radial and time coordinates
exchanged. A class of solutions, with spherical
subspaces, correspond
directly to the interior of black $p$-brane solutions. Solutions with
a non-trivial Neveu-Schwarz (NS) two-form field had been considered
previously by various authors~\cite{axion,be_fo}. Other authors have also
subsequently considered solutions with non-trivial RR forms. One of our
initial examples was presented at almost the same time in a paper by
Kaloper~\cite{kal}. A later paper by L\"u {\it et al.} gave a
further, broad class of solutions~\cite{lu}. The singularity-free cosmologies
arising inside black holes were first discussed by Behrndt and
F\"orste~\cite{be_fo} and then with RR fields by Poppe and
Schwager~\cite{rudi}. That these solutions really corresponded to the
interiors of black $p$-brane solutions was stressed recently by Larsen
and Wilczek~\cite{LW}, who also gave some further examples and
commented on the relation to D--branes. 

The purpose of the present paper is to show that the presence of non-trivial
form fields can actually stabilize the vacua of the dilaton and the
moduli that arise in the compactification of type II string theory or
M--theory. We will work within the context of the cosmological solutions that
we recently introduced in~\cite{letter,paper}, since understanding 
the stabilization of the dilaton and moduli is particularly physically
relevant in cosmology. However the mechanism is
more widely applicable. We give a simple argument as to how an
effective potential for both the dilaton and the moduli arises, and show
that this potential can have stable minima. The value of the fields at
the minimum is controlled by the form field charges. We then discuss, within
the context of type II and M-theory, the mechanism by which our
cosmological solutions are attracted to, and at late time stabilized
at, a finite minimum of the potential.

In the case of type II string theory, we find that 
the effect of exciting non-trivial
form fields is to introduce an effective potential for both the
dilaton and the geometrical moduli. Generically, 
the potential has flat directions or, even if there are no flat
directions, it may not have a minimum. Thus,
typically, not all the moduli are stabilized. However, in general, there
will be other contribution to the effective potential. In particular,
Tseytlin and Vafa~\cite{va_tsy} have argued in the cosmological
context that including a gas of string matter in toroidal
compactifications leads to an effective non-perturbative potential
for the geometric moduli, which tends to stabilize them
near the string scale. This non-perturbative potential is independent,
however, of the dilaton. Using this mechanism, we show in detail how
including non-trivial NS and RR form fields can then lead to a complete
stabilization of the dilaton. 

In the case of eleven-dimensional supergravity coming from M--theory,
the dilaton appears geometrically. Compactifying one direction on a circle
(or $S_1/Z_2$ orbifold), connects M--theory to type IIA string theory (or
heterotic string theory), where the dilaton is related to the radius
of the compact direction. We show that, in general, it is
possible to stabilize the moduli of several eleven-dimensional
compactifications simply by using form-fields. We also briefly
investigate the possibility of stabilizing those
configurations where the theory contains a dilaton. 

The layout of the paper is as follows. In section two, we explain the
mechanism by which non-trivial form fields can produce an effective
potential, by considering a simple toy example. 
In order to place this idea in a cosmological context, in section three
we summarize the structure of the cosmological solutions found
in~\cite{letter,paper}. Using spacetimes of this form, where the
internal space is a product of spheres and tori, we show explicitly
how an effective potential for the dilaton and geometrical moduli
describing the radii of the internal spaces can arise. In section four, we
show that, using a non-perturbative mechanism to stabilize the
geometrical moduli, the dilaton vacuum can be fixed by exciting a pair
of form fields. The fifth section is a discussion of how, in the
context of compactified supergravity, we are able to circumvent the
usual scaling argument that the dilaton cannot be stabilized. In the
sixth section, we discuss examples where the vacua of the geometrical
moduli arising from compactifying $M$--theory are fixed by this
mechanism. We briefly present our conclusions in section seven. 

%%%%%%%%%%%%%%%%%%%%%%%%%%%%%%%

\section{A simple example}

To understand how a non-trivial form field can lead to an effective
moduli potential, consider a simple model of a scalar field coupled
to an electromagnetic field strength in four dimensions, which mimics
the dilaton coupling in ten-dimensional supergravity, 
\be
  S = \int d^4x\, \left[- \frac{1}{4}e^{2\f}F^2  
    - \frac{1}{2}(\partial\f )^2 \right]\; .
    \label{4daction}
\ee 
As in the ten-dimensional theory the action has a symmetry under
$\f\rightarrow\f+\ln\l$ together with a rescaling of the gauge potential
$A_\m\rightarrow \l^{-1}A_\m$. Classically, the scalar field is the
massless Goldstone boson of this symmetry and so it would appear that
it has a flat effective potential. On the other hand the $\f$ equation
of motion reads
\be
  \partial^2\f = \frac{1}{2}e^{2\f}F^2 \; ,
\label{scalar}
\ee
so that $F$ could clearly supply an effective potential for the
scalar field, though naively it would appear not to have a minimum. 
Suppose that the $x$- and $y$-directions
are compactified so that the spacetime is a two dimensional Minkowski
space cross an ``internal space'' torus $T_2$.
The gauge field equation of motion and
Bianchi identity can be written as
\be
  d \left( * e^{2\f}F\right) = 0 \; , \qquad dF = 0 \; ,
\ee
respectively. Since there are no electric or magnetic charges
present, the flux of electric and magnetic field across the $x-y$
plane must be conserved, so we have the conserved charges 
\be
  e = \int_{T_2} * e^{2\f}F \; , \qquad g = \int_{T_2} F \; .
\label{em_charges}
\ee
The simplest configuration with these charges is a uniform electric
and magnetic field pointing along the $z$-direction, with the magnitudes
\be
  E_z = e^{-2\f} \frac{e}{A} \; , \qquad B_z = \frac{g}{A} \; ,
\label{EzBz}
\ee
where $A$ is the area of the torus. Substituting into
eq.~\refs{scalar}, we find  
\be
  \partial^2\f = 2 e^{2\f} F^2 
      = e^{2\f} \left( B_z^2 - E_z^2 \right)
      = \frac{g^2}{A^2} e^{2\f} - \frac{e^2}{A^2} e^{-2\f} 
      = \frac{dV_{\rm eff}}{d\f} \; .
\ee
where we define an effective potential for $\f$ of the form 
\be
  V_{\rm eff} = \frac{g^2}{2A^2} e^{2\f} + \frac{e^2}{2A^2} e^{-2\f} \; .
\ee
Note that, because of the factor of $\exp(2\f)$ which enters the
expression for the conserved electric charge~\refs{em_charges}, the
contribution to the effective potential from the electric field has the
opposite dependence on $\f$ from what might be expected from the form
of the action~\refs{4daction}. Clearly, the potential has a stable
minimum at $<\f >=\frac{1}{4}\ln\left(e^2/g^2\right)$. Remarkably,
exciting an electric and a magnetic field has not only provided an
effective potential, but, when both are present, has apparently
stabilized the scalar. The potential is, of course, a little naive. One
might wonder whether the electric and magnetic fields might not vary
with time in a way which removes this apparent stability. However the
charges~\refs{em_charges} must always be conserved, so, if the fields
do not depend on the compact directions, the only solution for $E_z$
and $B_z$ is that given by eqs.~\refs{EzBz}. From a compactification
point of view, allowing dependence on the $(x,y)$-directions corresponds to
exciting massive Kaluza-Klein modes, which would generally be expected
to raise rather than lower the vacuum energy. In fact, for a fixed
charge, the minimum contribution to the effective potential comes from
the case where the electric and magnetic fields are uniform. 
It is natural to ask how have we avoided the flat direction
implied by the scaling symmetry of the original
action~\refs{4daction}. The point is that the scaling symmetry does
{\em not} preserve the conserved charges~\refs{em_charges}. Rather, we
find $e\rightarrow \l e$ and $g\rightarrow \l^{-1}g$. Thus,
although making a scaling does provide a way of generating new
solutions, dynamically the new solution can never be reached from the
old one, since this would imply a violation of charge conservation. We
note that there is still remnant of the scaling symmetry in the
arbitrariness of the amount of electric and magnetic charge in the
solution. However, once these charges are chosen, the value of the
scalar field is stable and fixed. 

In summary, there are two parts to the stabilization. Without any electric or
magnetic field the theory has a scaling symmetry which implies that
the scalar field is always massless. Turning on one field, with its
corresponding conserved charge, breaks the symmetry and introduces an
effective potential for the scalar. However the potential has no
stable minimum. Only if the second field is excited do we get a
potential which stabilizes the scalar. Here the minimum arose from
balancing the magnetic and electric contributions of a single gauge
field. However, it is equally possible to produce the
stabilization using two different gauge fields. For instance, if the
Lagrangian has two terms $e^{2\f}F^2+e^{-2\f}F'^2$, two magnetic
fields can produce a minimum because of the different coupling each
term has to the scalar field. Such a theory would be analogous to
stabilizing between the NS and RR terms in type II theories.

With this simple example in mind, we can now see how this stabilization 
would work in a compactification of ten- or eleven-dimensional
supergravity. As usual, we take spacetime to be a product of a
four-dimensional space $M_4$ with some compact internal space $K$. (In
our examples $M_4$ will be a Robertson-Walker cosmology, while $K$ is a 
product of spheres and tori.) Considered as a theory in four
dimensions, in addition to the dilaton there are moduli $\a_i$
describing the internal
space (for instance, the radii of the internal spheres and tori in our
examples). These degrees of freedom appear in the metric for
the internal space and so will also couple to the form-field $F^2$
terms in the supergravity action. In our simple examples they appear
as exponential coefficients, exactly analogous to the dilaton
coupling. This coupling implies that, as we will see, exciting
non-trivial form fields can provide an effective potential for
the metric moduli of the internal space as well as for the dilaton.

What form field orientations should we take? We start by
noting that, for all the examples we consider, the
contribution from possible Chern-Simons in the supergravity theory are
zero. Thus we may effectively drop these terms from the action. 
Suppose there is a $\d$-form
potential which appears in the action as $\exp(-a\f)F_\d^2$. We can
again form conserved electric and magnetic charges
\be
  e_\d = \int_{\Sigma_{\tilde{\d}+1}} * e^{-a\f} F_\d \; , \qquad
  g_{\tilde{\d}} = \int_{\Sigma_{\d+1}} F_\d \; ,
\ee
where $D$ is the total spacetime dimension and $\tilde{\d}=D-\d -2$. The
integrals are over compact subspaces $\Sigma_{\tilde{\d} +1}$ and
$\Sigma_{\d+1}$ which, by analogy to the four-dimensional case, lie only in
the internal space $K$. Thus, we are interested in the cases where either
$F_\d$ or $* F_\d$ lies only in $K$. It is immediately clear that we
can only have an electric charge if $\d>2$, since otherwise $*
F_\d$ must lie partly in $M_4$. This implies, for instance, that we
cannot excite an electric charge in heterotic string theory, since there
are no massless form fields with $\d>2$. Likewise, though less
relevantly, for a magnetic charge $\d<D-4$. In the effective
four-dimensional theory, a given ten-dimensional form field appears as
of four-dimensional form of varying degree, depending on how many
components of the form span the internal space. Our condition on
$F_\d$ implies that in four-dimensions we have either a 0-form
(magnetic case) or a four-form (electric case). Such dimensional
reductions, though with only magnetic charges, appear in the derivation of
``massive'' supergravity theories given in~\cite{massSUGRA}. 
In either case, the
field has no four-dimensional dynamics and simply provides an
effective potential for the moduli and dilaton. In fact, in both
cases, since we assume that the dilaton depends only on the 
external coordinates, the equations of motion for the form field
reduce to  
\be
  d * G = d G =0 \; ,
\ee
where the forms, exterior derivative and Hodge star are now restricted
to the internal subspace $K$. Here $G$ is the projection of $F_\d$ onto
$K$ in the magnetic case, while in the electric case it is the
projection of $*F_\d$. The conditions imply that $G$ is
harmonic in $K$. For each independent solution, we can fix
either the electric or magnetic charge. In our examples, we will often
refer to these two cases as ``fundamental'' (for an electric charge)
and ``solitonic'' (for a magnetic charge) by analogy with the fields
surrounding fundamental and solitonic $p$-branes. With this formalism
in mind, we now give a detailed analysis of dilaton and moduli vacuum
stabilization within the context of cosmological type II and M--theory
solutions. 

\section{Cosmological framework}

We are interested in cosmological solutions of a low energy action,
which can describe the bosonic modes of type II or eleven-dimensional
supergravity, written in the Einstein frame,
\be
 S=\int d^Dx\;\sqrt{-g}\left[ R-\frac{4}{D-2}(\partial\f )^2-\sum_r
    \frac{1}{2(\d_r+1)!}e^{-a(\d_r)\f}F_r^2-\L_8 e^{-a_{\L_8} \f}\right]
    \label{action}
\ee
Here ${g}_{MN}$ is the $D$--dimensional metric,
$\f$ is the dilaton and $F_r=dA_r$ are $\d_r$--forms.
The forms encompass the NS two--form as well as the RR forms
present in type II theories. We have also included a cosmological constant
$\L_8$ which appears in the massive extension of type IIA
supergravity~\cite{romans} and can be interpreted as a
RR 9--form coupling to 8 branes~\cite{bergs_8brane}.
Here we adopt the Ansatz that none of our solutions include
contributions from Chern-Simons terms, and so, such terms can
therefore be dropped from the action.
The various types of forms are distinguished from each other by the
dilaton couplings $a(\d_r )$ which are given by
\be
 a(\d_r) = \left\{ \ba{cll} \frac{8}{D-2}&{\rm NS}&{\rm 2-form}\\
                      \frac{4\d_r -2(D-2)}{D-2}&{\rm RR}&\d_r{\rm -form}\\
                   \ea\right. \label{p_rr}
\ee
and
\be
 a_{\L_8} = -\frac{2D}{D-2}\; .  \label{p_lambda}
\ee
Note that the couplings for the NS form and the RR forms have opposite
signs.
The above action can account for a wide range of cosmological solutions
in type II theories and M--theory, and a large class of such
solutions has been constructed in~\cite{paper}. Since it is this class
which we will use to illustrate our main idea, let us briefly review some
of its properties.

Our Ansatz for the metric is characterized by a split of the
total $D$--dimensional space into $n$ maximally symmetric $d_i$
dimensional spatial subspaces with scale factors $\bar{\a}_i$, $i=0,...,n-1$.
The corresponding metric reads
\be
  ds^2 = -\bar{N}^2(\t )d\t^2
     +\sum_{i=0}^{n-1}e^{2\bar{\a}_i(\t )}d\O_{K_i}^2\; ,
 \label{c_metric}
\ee
where $d\O_{K_i}^2$ is the metric of a $d_i$ dimensional space with constant
curvature $K_i=-1,0$ or $+1$. The dilaton should also depend only on
time, $\f =\f (\t )$. We have in mind that three of the
spatial directions should be identified with the spatial part of the
observed universe. Typically, these three directions correspond to one
subspace to ensure homogeneity as well as isotropy. One might, however,
also allow for a further split up of this three dimensional part, if
the resulting anisotropy disappears asymptotically in time. The other
directions should be viewed as a compact internal space and the
corresponding scale factors are interpreted as moduli.

The structure of eq.~\refs{c_metric} allows two different types
of Ans\"atze for the forms, which we call ``elementary'' and ``solitonic''.
This terminology is motivated by a close analogy to p--brane solutions
which has been explained in ref.~\cite{letter,paper}. The nonvanishing
components of their field strengths are given by
\begin{itemize}
 \item elementary~: if $\sum_i d_i=\d_r$ for some of the spatial subspaces $i$
                    we may set
 \be
  (F_r)_{0\m_1...\m_{\d_r}} = A_r(\bar{\a} )\, f_r'(\t )\,\e_{\m_1...\m_{\d_r}}\; ,
  \quad A_r(\bar{\a} ) = e^{ -2\sum_i d_i\bar{\a}_i } \label{elementary}
 \ee
 where $\m_1...\m_{\d_r}$ refer to the coordinates of these subspaces,
 while $\e_{\m_1...\m_{\d_r}}$ is a totally antisymmetric tensor
 density spanning these subspaces,
 $f_r(\t )$ is an arbitrary function to be fixed by the form field
 equation of motion, and the prime denotes the derivative with
 respect to $\t$.  Note that the sum over $i$ in the exponent runs only over
 those subspaces which are spanned by the form. 
 The ``electric'' configurations discussed in the
 last section correspond to an elementary form which spans all the
 external subspaces.
 \item solitonic~: if $\sum_i d_i=\d_r+1$ for some of the spatial subspaces $i$
                   we may allow for
 \be
  (F_r)_{\m_1...\m_{\d_r+1}} = B_r(\bar{\a} )\; w_r\; \e_{\m_1...\m_{\d_r+1}}\;
 ,\quad
  B_r(\bar{\a} ) = e^{ -2\sum_i d_i\bar{\a}_i } \label{solitonic}
 \ee
 where $\m_1...\m_{\d_r+1}$ refer to the coordinates of these subspaces and
 $w_r$ is an arbitrary constant. As for the elementary Ansatz, the sum
 over $i$ in the exponent runs over the subspaces spanned by the
 form. It is easy to check that this Ansatz already solves the form
 equation of motion. The ``magnetic'' configurations discussed in the
 last section correspond to a solitonic form which spans no part of
 the external subspaces.
\end{itemize}

\vspace{0.4cm}

{}From the form of the action~\refs{action}, it is clear that the two Ans\"atze
for the forms~\refs{elementary} and \refs{solitonic} generate an effective
potential for the scale
factors as well as for the dilaton. Also, terms resulting from curved
subspaces can be incorporated into this potential. In ref.~\cite{paper}
this has been made precise by deriving an effective action for the vector
$\bar{\bal} = (\bar{\a}_I)=(\bar{\a}_i ,\f )$ 
\be
 {\cal L} = \frac{1}{2}E\bar{\bal} '^T\bar{G}\bar{\bal} '-E^{-1}U
  \label{dan_lag}\; .
\ee
where the metric $\bar{G}_{IJ}$ on the $\bar{\bal}$ space is defined by
\bea
 \bar{G}_{ij}&=&2(d_i\d_{ij}-d_id_j)\nn \\
 \bar{G}_{in}&=&\bar{G}_{ni}=0 \label{G}\\
 \bar{G}_{nn}&=&\frac{8}{D-2}\; .\nn 
\eea
and
\be
 E=\frac{1}{\bar{N}}\exp (\bar{\bd}\cdot\bar{\bal}) \label{E}
\ee
with the dimension vector $\bar{\bd}=(d_i,0)$. The effective potential
$U$ can be written as a sum
\be
 U = \frac{1}{2}\sum_{r=1}^{m}u_r^2\exp (\bar{\bq}_r \cdot\bar{\bal} )
     \label{U}
\ee
over all elementary and solitonic form configurations as well as all
curvature terms. The type of a certain term $r$ is specified by the
vector $\bar{\bq}_r$. For an elementary $\d$--form it is given
by
\be
 \bar{\bq}^{\rm (el)} = (2\e_id_i,a(\d ))\; ,\quad \e_i=0,1\; ,
 \quad \d =\sum_{i=0}^{n-1}\e_id_i \label{q_el}
\ee 
with $\e_i=1$ if the form is nonvanishing in the subspace $i$ and
$\e_i=0$ otherwise. The dilaton couplings $a(\d )$ have been defined in
eq.~\refs{p_rr}. For a solitonic $\d$--form it reads
\be 
 \bar{\bq}^{\rm (sol)} = (2\tilde{\e}_id_i,-a(\d ))\; ,\quad
  \tilde{\e}_i\equiv 1-\e_i
 =0,1\; ,\quad \tilde{\d}\equiv D-2-\d =\sum_{i=0}^{n-1}\tilde{\e}_id_i
 \label{q_sol}
\ee
with $\tilde{\e}_i=1$ if the form vanishes in the subspace $i$ and
$\tilde{\e}_i=0$ otherwise. In both cases, the constant $u_r^2$ in
potential~\refs{U} is a positive integration constant, proportional to
the square of the conserved electric of magnetic form-field
charge. Finally, curvature in the $k$th subspace leads to a potential
term characterized by 
\be 
 \bar{\bq}^{\rm (curv)}_k = (2(d_i-\d_{ik}),0)\; . \label{curv}
\ee
In this case the constant $u_r^2$ is determined by the curvature,
$u_r^2=-2K_k$, and can be of either sign. 

The dynamical properties of models specified by Lagrangian~\refs{dan_lag}
and potential~\refs{U} have been studied at length in refs.~\cite{paper}.
In particular, the general solution for models with only one term in the
potential and the solution for those models related to Toda theory have
been found. Here, though, we concentrate on the question of dilaton and moduli
vacuum stabilization.

\vspace{0.4cm}

So far, we have treated all scale factors on the same footing. Physically,
however, it is useful to distinguish scale factors of the observable
universe from internal moduli arising from compactification, and to
transform to a new basis in which these two types of fields decouple.

To do this, let us split up the $\bar{\bal}$ space into an external observable
universe part, an internal moduli part and the dilaton as
$\bar{\bal}=(\bar{\bal}^{(e)}_\b,\bar{\bal}^{(i)}_b,\f )$. Note that we
use indices $\b ,\g ,...$ to specify the scale factor(s) of the universe
and indices $b,c,...$ to specify the moduli. Of course, we have a split
into $3+1$ external dimensions and either $6$ (string theory) or $7$
(M--theory) internal dimensions in mind; that is
$D^{(e)}\equiv\sum_\b d_\b =3$ and $D^{(i)}\equiv\sum_b d_b =6$ or $7$.
All other vectors will be split correspondingly, for example
$\bar{\bq}=(\bar{\bq}^{(e)},\bar{\bq}^{(i)},\bar{q})$. 
As it stands, Lagrangian~\refs{dan_lag} mixes the external and the
internal spaces since the metric $\bar{G}_{ij}$, eq.~\refs{G}, is
completely off--diagonal. To decouple these spaces we should, instead
of action~\refs{action}, consider its dimensional reduction 
$\sqrt{-g_{10}}R_{10}\rightarrow \sqrt{-g_4}R_4\; +$ moduli. In order
to do this reduction one has to perform a Weyl rotation on the external
metric. Within our framework this Weyl rotation can be simply described
by a linear transformation $\bal =P^{-1}\bar{\bal}$ to a new basis $\bal$
and a corresponding transformation of the gauge parameter $N$. In the basis
$\bal$ the new metric $G=P^T\bar{G}P$ is block diagonal in the internal and
external parts. It turns out that the transformation is given by
\bea
 \bar{\a}_\b &=& \a_\b - \frac{1}{D^{(e)}-1}\sum_bd_b\a_b \nn \\
 \bar{\a}_b &=& \a_b \label{trafo} \\
 \bar{\f} &=& \f \; .\nn \\
 \bar{N} &=& e^{-\sum_cd_c\a_c/(D^{(e)}-1)}N \nn
\eea
Comoving time $t$ in the new frame is defined by setting $N=1$.
The new metric $G$ is explicitly given by
\be
 G = \left(\ba{ccc} G^{(e)}&0&0\\
                    0&G^{(i)}&0\\
                    0&0&\frac{8}{D-2}
     \ea\right)
\ee
with
\bea
 G^{(e)}_{\b\g} &=& 2(d_\b\d_{\b\g}-d_\b d_\g )
 \label{G_e} \\
 G^{(i)}_{bc} &=& 2\left( d_b\d_{bc}+\frac{1}{D^{(e)}-1}d_b d_c\right)
\eea
Note that the external part of the metric $G^{(e)}$ is unchanged, as it
should be. Correspondingly, one should compute the vectors
$\bq = P^T\bar{\bq}$ in the new basis. This can be worked out in a
straightforward way, and eq.~\refs{trafo} shows that only the internal part
of these vectors changes. We find that elementary and solitonic forms are
now characterized by
\bea
 \bq^{\rm (el)} &=& \left(2\e_\b d_\b ,2\left(\e_b-\frac{\d^{(e)}}{D^{(e)}-1}
                \right)d_b ,a(\d )\right) \label{el_new}\\   
 \bq^{\rm (sol)} &=& \left(2\tilde{\e}_\b d_\b ,2\left(\tilde{\e}_b-
                 \frac{\tilde{\d}^{(e)}}{D^{(e)}-1}\right)d_b ,-a(\d )\right)
                 \label{sol_new}
\eea
where $\d^{(e)}=\sum_\b\e_\b d_\b$ is the ``overlap'' of an elementary
form with the external space and $\tilde{\d}^{(e)}=\sum_\b\tilde{\e}_\b d_\b$
is the complementary quantity for a solitonic form. The vectors specifying
curvature in the external and internal space are given by
\bea
 \bq^{\rm (curv)}_\b &=& (2(d_\g -\d_{\b\g}),0 ,0) \label{curve_new}\\
 \bq^{\rm (curv)}_b &=& \left( 2d_\g ,2\left( -\frac{1}{D^{(e)}-1}d_c-\d_{cb}
                    \right) ,0\right) \; .
 \label{curvi_new}
\eea 
The quantity $E$ defined in eq.~\refs{E} should be rewritten in terms
of the transformed quantities as
\be
  E=\frac{1}{N}\exp (\bd\cdot\bal )\; .\label{E_new}
\ee
{}From transformation~\refs{trafo} we read off the new dimension vector
\be
 \bd = \left(d_\b,0,0\right)\; .
 \label{d_new}
\ee
Note that the internal components of this vector vanish. This could have been
anticipated from the fact that we are actually performing a dimensional
reduction, so $E$ should depend on the reduced metric only.

\vspace{0.4cm}

We can now rewrite Lagrangian~\refs{dan_lag} in terms of the new, unbarred
quantities as
\be
 {\cal L} = \frac{1}{2}E{\bal^T} 'G\bal ' -E^{-1}U \\
\ee
where, as before, $U=\frac{1}{2}\sum_{r=1}^{m}u_r^2\exp (\bq_r\cdot\bal )$.
Since the new metric $G$ is block diagonal, we can separate the equations of
motion into an external, an internal and a dilaton part as
\bea
 \frac{d}{d\t}\left( EG^{(e)}{\bal^{(e)}} '\right)+
   E^{-1}\frac{\partial U}{\partial\bal^{(e)}}&=&0 \label{ale_eom} \\ 
  \frac{d}{d\t}\left( EG^{(i)}{\bal^{(i)}} '\right)+
   E^{-1}\frac{\partial U}{\partial\bal^{(i)}}
  &=&0 \label{ali_eom} \\ 
  \frac{d}{d\t}\left( E\frac{8}{D-2}\f '\right)+
   E^{-1}\frac{\partial U}{\partial\f}
  &=&0 \label{phi_eom} \\ 
 \frac{1}{2}E{{\bal^{(e)}} '}^TG^{(e)}{\bal^{(e)}} '+
 \frac{1}{2}E{{\bal^{(i)}} '}^TG^{(i)}{\bal^{(i)}} '+
 E^{-1}U &=& 0\; . \label{N_eom}
\eea
The last equation is a constraint which arises as the equation of
motion for the gauge parameter $N$. It is useful to rewrite the equations of
motion for the moduli and the dilaton in terms of the comoving time $t$
defined by $N=1$. Defining a modified potential $V$ by
\be
 V=\exp (-2\bd^{(e)}\cdot\bal^{(e)})U \label{V_def}
\ee
we get from eq.~\refs{ali_eom}, \refs{phi_eom}, \refs{E_new} and \refs{d_new}
that
\bea
 G^{(i)}\ddot{\bal}^{(i)}+(\bd^{(e)}\cdot\dot{\bal}^{(e)})
  G^{(i)}\dot{\bal}^{(i)}+\frac{\partial V}{\partial\bal^{(i)}} &=& 0
  \label{moduli}\\
 \ddot{\f}+(\bd^{(e)}\cdot\dot{\bal}^{(e)})\dot{\f}+\frac{\partial V}
  {\partial\f} &=& 0 \label{phi}
\eea
where the dot denotes the derivative with respect to $t$. The potential $V$
is explicitly given by
\be
 V = \frac{1}{2}\sum_{r=1}^{m}u_r^2\exp ((\bq_r^{(e)}-2\bd^{(e)})
     \cdot\bal^{(e)})\exp (\bq_r^{(i)}\cdot\bal^{(i)})\exp (q_r\f )\; . 
 \label{V}
\ee
The potential $V$ can be interpreted as the effective moduli and dilaton
potential in the dimensionally reduced 4--dimensional external space action.
Correspondingly, the above equations are exactly those of scalar fields with
a potential $V$ in an expanding universe. The potential is provided by
the forms and the curvature terms. There is, however, one difference from
the ordinary case. Unlike a usual scalar field potential, $V$ can also
depend on the external scale factors $\bal^{(e)}$ so that, in general, its
shape changes due to the evolution of the universe. Let us analyze this in
detail. The terms in potential~\refs{V} with $\bq^{(e)}_r= 2\bd^{(e)}$
have no dependence on external scale factors and can be viewed as the ``true''
potential. A comparison of the $\bq$ vectors in eq.~\refs{el_new},
\refs{sol_new}, \refs{curve_new}, \refs{curvi_new} with $\bd$ in
eq.~\refs{d_new} shows that the entries of $\bq^{(e)}_r$ are always smaller
or equal to $2\bd^{(e)}$. Therefore, all other terms with
$\bq^{(e)}_r\neq 2\bd^{(e)}$ are suppressed at late time if the universe
expands. In terms of the reduced four-dimensional theory, these
suppressed terms correspond to exciting four-dimensional matter in the
form of 0-, 1- or 2-form potentials. The suppression implies that, at
late time, we have 
\be
 V\simeq \frac{1}{2}\sum_{\bq_r^{(e)}=2\bd^{(e)}}u_r^2
     \exp (\bq_r^{(i)}\cdot\bal^{(i)})\exp (q_r\f )\; ,
\ee
where, as indicated, the sum now runs over all terms with
$\bq^{(e)}_r= 2\bd^{(e)}$. Which forms and curvature terms can actually
contribute to this late time potential? Eq.~\refs{el_new} shows that
an elementary form should occupy the whole external space to meet this
requirement. As such, it corresponds exactly to the ``electric''
configuration discussed in the previous section. 
If the external space is 3--dimensional, this can be done
with the 3--form of type IIA or M--theory. On the other hand, from
eq.~\refs{sol_new}, a solitonic form should have nonvanishing components in
the internal space only, and corresponds to the ``magnetic''
configuration discussed in the previous section.  
Finally, a curvature term contributes to the
late time potential if it describes a curved internal space, as
can be seen from eqs.~\refs{curve_new} and \refs{curvi_new}.

We see that there are a number of possible sources for the asymptotic
potential at late time within our framework. It is conceivable that this
can be used to stabilize the dilaton and/or the moduli at a finite minimum of
$V$. We will now address this question in detail, distinguishing two
cases. As the first case, we assume that the dilaton is not the modulus of
any compactification but acts as the string coupling constant only. This
is the pure type II (string) theory point of view. As the second case, we
assume that the dilaton is on the same footing as the moduli; that is, it
is a modulus itself (related to the compactification from $D=11$ to
$D=10$). This is the M--theory point of view.

\section{Stabilizing the dilaton in type II}

Let us consider the first case, when $\f$ is not a geometrical modulus. In
general, the potential provided by the forms is not sufficient to fix
all the moduli and the dilaton. However, we would like to
show that it can fix the dilaton vacuum 
once the moduli vacua have been fixed. To fix the moduli, we assume the
existence of a nonperturbative potential $V_{\rm np}(\bal^{(i)})$, which
depends on the moduli only. This is added to $V$,
\be
 V_{\rm T} = V+V_{\rm np}(\bal^{(i)})\; , \label{V_t}
\ee
and should have a minimum to which a sufficiently
large set of trajectories is attracted at late time.
A concrete realization is, for example, provided by the
mechanism discussed by Tseytlin and Vafa~\cite{va_tsy}. They have shown
that the inclusion of string matter both in the form of momentum and
winding modes around a compact direction can stabilize a modulus if
both types of matter fail to annihilate. The momentum modes prevent
the compact direction from collapsing and the winding modes around that
direction prevent expansion. Clearly, since we assume that the dilaton is
not a geometrical modulus and therefore does not correspond to a
compact direction, such a mechanism cannot be invoked to provide
dilaton stabilization. 

We now analyze under what conditions the dilaton can be stabilized.
At early times, the moduli, as well as the dilaton, will be displaced from
their minima and, finally, oscillate around them. Since we are mainly
interested in the vacuum of the dilaton in the present epoch, we will not
address this early period, but rather attempt to find a late time
asymptotic solution.

First, write out the total late time potential $V_{\rm T}$ as
\be
 V_{\rm T} \simeq \frac{1}{2}\sum_{\bq_r^{(e)}=2\bd^{(e)}}u_r^2
       \exp (\bq_r^{(i)}\cdot\bal^{(i)})\exp (q_r\f )+
       V_{\rm np}(\bal^{(i)})\; . \label{V0}
\ee
The nonperturbative potential $V_{\rm np}$ has been included to stabilize
the moduli and we have assumed that it is of the appropriate form to do so.
We have, however, to guarantee that a constant moduli solution survives
if the whole potential $V_{\rm T}$ is taken into account. Let us make the
consistency assumption that the dilaton is fixed 
(to be verified later). The total potential $V_{\rm T}$ should then
still have a minimum with a sufficiently large basin of attraction. This
is, for example, true if $V_{\rm np}\rightarrow\infty$ for
$|\bal^{(i)}|\rightarrow\infty$ (a requirement which is fulfilled by the
mechanism of Tseytlin and Vafa) and the form and curvature potential is
bounded from below (This is true for all possible sources except for positive
curvature subspaces. These diverge for small scale factors and have to be
balanced by a positive form contribution to fulfill the
requirement). These conditions allow us to 
assume the existence of a well defined moduli minimum
$<\bal^{(i)}>$ for $V_{\rm T}$. Then $\bal^{(i)}=<\bal^{(i)}>$ fulfills
the moduli equation of motion~\refs{moduli} and potential~\refs{V0} turns into
\be
 V_{\rm T} \simeq \frac{1}{2}\sum_{\bq_r^{(e)}=2\bd^{(e)}}\tilde{u}_r^2
       \exp (q_r\f )+\L_{\rm np}\; ,  \label{V1}
\ee 
where
\bea
 \tilde{u}_r^2&=&u_r^2\exp (\bq_r^{(i)}\cdot<\bal^{(i)}>) \label{u_tilde}\\
 \L_{\rm np}&=&V_{\rm np}(<\bal^{(i)}>)\; .
\eea
Note that $\L_{\rm np}$ is the contribution to the cosmological constant
which results from the nonperturbative moduli potential. Its actual value
depends on the specific mechanism which has been invoked to
create $V_{\rm np}$.

In order for the dilaton to have a minimum, the sum in~\refs{V1} 
should contain at least two terms with opposite sign of the dilaton
coupling $q_r$. The analog of the simple toy example given in section
two, would be to excite a solitonic orientation and a fundamental
orientation of the same form, since, from eqs.~\refs{el_new},
\refs{sol_new}, we see the two orientations do have different signs in
the dilaton coupling. Note, however, that the elementary part has to
cover the {\em full} external space in order to get a ``real'' potential
term which is not suppressed for a large observable universe. Though the
NS 2--form could provide both a solitonic and an elementary Ansatz, the
elementary part does not fully cover a $3+1$--dimensional external space.
Therefore the corresponding potential term drops as the universe expands
and the dilaton cannot be stabilized. However, an elementary RR 3--form
fits into a $3+1$ dimensional external space. With an additional solitonic
3--form entirely in the internal space the potential indeed has a stable
minimum. The problem is that as a result the
Chern-Simons contribution to the IIA equations of motion does not
vanish. This takes us outside our Ansatz, and, for this reason, while such
a configuration may provide a way of stabilizing the dilaton, we will
ignore this possibility from here on. 

The remaining possibility with opposite sign dilaton couplings is
to turn on a solitonic NS 2--form and a solitonic RR form in the
internal space, as can be seen from the dilaton
couplings~\refs{p_rr}. Note that, in order to have the opposite sign of
their dilaton couplings, it is crucial to have a RR form turned on in
addition to the NS form. Under this condition we indeed have a
solution $\f = <\f >$ for the dilaton equation of motion~\refs{phi},
where $<\f >$ is the minimum of ~\refs{V1}. The value of $<\f >$ will
consequently be controlled by the strengths of the form fields given by the
appropriate $u_r$ parameters and the vacuum values of the moduli. To conclude,
under very mild restrictions on the forms and the structure of the
nonperturbative potential, we have found that the dilaton approaches a
constant value $<\f >$ at late time. With the fixed dilaton, the late
time potential~\refs{V1} turns into a pure cosmological constant
\be
 \L = \L_{\rm f}+\L_{\rm np}\; ,
\ee
where
\be
 \L_{\rm f} = \frac{1}{2}\sum_{\bq_r^{(e)}=2\bd^{(e)}}\tilde{u}_r^2
       \exp (q_r<\f > )
\ee
is the contribution to the cosmological constant arising from the forms and
curvature terms. The only negative contribution to $\L_{\rm f}$ arises
{}from internal subspaces with positive curvature. If they are absent,
$\L_{\rm f}$ is positive, otherwise it can be of either sign or it can
vanish.

\vspace{0.4cm}

As the last step in constructing a consistent late time solution, we should
analyze the behaviour of the external scale factors $\bal^{(e)}$. To do so,
we need the effective potential
$U_{\rm T}=\exp (2\bd^{(e)}\cdot\bal^{(e)})V_{\rm T}$
which arises after fixing the moduli and the dilaton. Inserting this into 
eq.~\refs{N_eom} then determines the evolution of the external
scale factors. From eq.~\refs{V} we have
\be
 U_{\rm T} = \L \exp (2\bd^{(e)}\cdot\bal^{(e)}) + \frac{1}{2}
       \sum_{\bq^{(e)}\neq 2\bd^{(e)}}\tilde{\tilde{u}}_r^2
       \exp (\bq ^{(e)}_r\cdot\bal^{(e)} ) \label{U_fix}
\ee
with $\tilde{\tilde{u}}_r^2=\tilde{u}_r^2\exp (q_r<\f > )$.
Note that the first term represents the cosmological constant which, as
discussed above, arises from the late time potential. In addition,
we have all those terms from eq.~\refs{V} with $\bq^{(e)}\neq 2\bd^{(e)}$
that decay at late time if the universe expands. Clearly, these terms
can be neglected at late time if the cosmological constant is positive,
$\L > 0$. Let us consider this case first.

For simplicity, assume that the external space is spatially isotropic; that is
$\bal^{(e)}=(\a_0 )$ and $\bd^{(e)} = (3)$. Then, from eq.~\refs{G_e}, we have
$G^{(e)}=(-12)$ and from eq.~\refs{E_new} it follows that
$E=\exp (3\a_0 )$ in the comoving gauge $N=1$. Inserting these quantities
into eq.~\refs{N_eom}, we find $\dot{\a }_0 = \sqrt{\L /6}$. This
corresponds to a de Sitter spacetime with inflationary expansion.

It might be possible to tune the two contributions to $\L$ such
that $\L =0$. In that case, if there is no other source of energy density,
the universe is static, $\dot{\bal}^{(e)}=0$.
Energy density for an expansion could
be provided by the other terms in eq.~\refs{U_fix} related to forms or
curvature terms which are ``nontrivial'' in the external space. In
terms of the reduced four-dimensional action, such terms correspond to
exciting form-field matter, or curving the spatial part of the
four-dimensional metric. It results
in a radiation--like expansion. Let us again consider the case of an
isotropic external space. Then the only possibility to have
$\bq^{(e)}\neq 2\bd^{(e)}$ is $\bq^{(e)}=(0)$. In the reduced theory
this corresponds to exciting the kinetic terms of 0-form, that is
scalar, matter. From eq.~\refs{U_fix}
we find $U=$ const and eq.~\refs{N_eom} can be readily solved to
give $\a_0 = \ln t/3 + c$, where $c$ is a constant related to $U$. As 
expected, this subluminal expansion with a 
power $1/3$ is characteristic for an expansion
driven by scalar field 
kinetic energy. If the external space is non-isotropic, one may
have nonzero vectors $\bq^{(e)}$ with $\bq^{(e)}\neq 2\bd^{(e)}$ and the
solution is more complicated. Its general form, if only one of those terms
appears in the potential~\refs{U_fix}, has been given in ref.~\cite{paper}.
In this case, the expansion is subluminal, radiation--like but the expansion
powers, though always smaller than one, may depart from the value $1/3$.
Alternatively, one could add radiation to the model which
one expects to arise from the decay of the coherent moduli and dilaton
oscillations at early times. This would yield a true radiation dominated
phase with an expansion power $1/2$.

\vspace{0.4cm}

To summarize, we have shown that, within our Ansatz, in order to
stabilize the dilaton vacuum it is essential to have a solitonic NS
2--form and a solitonic RR form both turned on in the internal
space. The value of the dilaton vacuum is controlled by the charges of
the form fields and the vacuum values of the moduli. Having additional forms
does not change this result. We emphasize, that the existence of RR forms
is crucial for this mechanism to work because of their different couplings
to the dilaton in action~\refs{action}. 

\vspace{0.4cm}

Let us illustrate the above mechanism with a concrete
$D=10$, type IIA example. We choose an external $1+3$--dimensional
space ($D^{(e)}=3$, $\bd ^{(e)}=(6)$) with scale factor
$\bal^{(e)}=(\a_0)$. The internal 6--dimensional space ($D^{(i)}=6$) is
split up as $3+2+1$ so that we are dealing with three moduli
$\bal^{(i)}=(\a_1,\a_2,\a_3)$. For simplicity, we take all spatial subspaces
to be flat. 

The internal space has been split in this particular way so as to place a
solitonic NS 2--form in the 3--dimensional subspace and a solitonic
RR 1--form in the 2--dimensional subspace. From eq.~\refs{sol_new}
we find that these two forms are described by the vectors
$\bq_{\rm NS}=(6,-9,-2,-1,-1)$ and $\bq_{\rm RR}=(6,-3,-6,-1,3/2)$.
Note that for both vectors
$\bq^{(e)}_{\rm NS}=\bq^{(e)}_{\rm RR}=2\bd^{(e)}=(6)$, so that they are
maximal on the external space and contribute to the late time potential.
Furthermore, we have the internal vectors $\bq^{(i)}_{\rm NS}=(-9,-2,-1)$,
$\bq^{(i)}_{\rm RR}=(-3,-6,-1)$ and the dilaton couplings $q_{\rm NS}=-1$,
$q_{\rm RR}=3/2$. Then, from eq.~\refs{V} we find the potential
\be
 V = \frac{1}{2}\left( u_{\rm NS}^2e^{-9\a_1-2\a_2-\a_3}e^{-\f}+
                       u_{\rm RR}^2e^{-3\a_1-6\a_2-\a_3}e^{3\f /2}\right)\; .
 \label{V_ex}
\ee
Clearly, this potential has a minimum in the dilaton direction.
As it stands, however, it drives the moduli to infinity
and their variation in time then also renders the dilaton minimum time
dependent. Therefore, we assume that the moduli $\a_1 , \a_2 , \a_3$
(but not the dilaton!) are stabilized by some nonperturbative potential
$V_{\rm np}(\bal^{(i)})$. (The possibility of stabilizing all fields,
the dilaton and the moduli, without invoking any nonperturbative
effects will be analyzed below). After a sufficiently long time,
oscillations are damped out and the moduli have settled down to
their minimum $\bal^{(i)}=<\bal^{(i)}>$. Then potential~\refs{V_ex}
turns into
\be
 V=\frac{1}{2}\left(\tilde{u}^2_{\rm NS}e^{-\f}+\tilde{u}^2_{\rm RR}
   e^{3\f /2}\right)\; . \label{V1_ex}
\ee
The constants $\tilde{u}^2_{\rm NS}$ and $\tilde{u}^2_{\rm RR}$ are 
defined as in eq.~\refs{u_tilde}. This potential has a dilaton
minimum at
\be
 <\f > =\frac{2}{5}\ln\left(\frac{2\tilde{u}^2_{\rm NS}}
       {3\tilde{u}^2_{\rm RR}}\right) \label{Vmin_ex}
\ee
with positive cosmological constant
\be
 \L_{\rm f} = \frac{5}{6}\left(\frac{2}{3}\right)^{-2/5}
              (\tilde{u}^2_{\rm NS})^{3/5}
              (\tilde{u}^2_{\rm RR})^{2/5}\; .
\ee
If the total cosmological constant $\L = \L_{\rm f}+\L_{\rm np}$ is
positive, the external space expands in a de Sitter phase. If $\L =0$
the external space is static. By adding radiation (for example from the decay
of the early time oscillations) to our model we can also get a radiation
dominated phase.

\section{A scaling argument}

It is useful, at this point, to present a more physical explanation of why
the dilaton vacuum can be determined, along the lines of the
discussion for the simple model given in section two. For 
clarity, we focus on the specific
example just discussed which will graphically illustrate our main point.
Any other solution can be analyzed in a similar manner. Consider the
action~\refs{action} restricted to this specific example. The relevant
fields are, in addition to the metric and dilaton, a NS 2--form $B_{\m\n}$
and a RR $\d$--form $A_{\m_1...\m_\d}$, each living in the internal space
only. We will restrict the form fields to be solitonic, in line with
the example, though we will set $\d =1$ only later.
We make the physical assumption that the observable space
continuously expands but that the compactified space, after a period of
contraction, becomes fixed. Let us scale the fields according to
\bea
 \f&\rightarrow&\f +s\ln \l \nn \\
 B_{\m\n} &\rightarrow&\l^{\frac{r}{2}+\frac{s}{2}}B_{\m\n} \label{form_scal}\\
 A_{\m_1...\m_\d}&\rightarrow&\l^{\frac{r}{2}+\left(\frac{\d -4}{2}\right)
                \frac{s}{2}}A_{\m_1...\m_\d}\nn \; .
\eea
Furthermore, we scale the $3+1$--dimensional part of the metric as
\be
 g_{\m\n}\rightarrow \l^{-r}g_{\m\n} \label{g_scal}
\ee
but hold the $6$--dimensional internal space metric fixed.
The action~\refs{action} is, up to an overall factor, invariant
under these transformations for arbitrary values of $r$, $s$ and, hence,
so are the equations of motion.

It follows from the invariance of the action under the Abelian gauge
transformation $B\rightarrow B +d\L_2$ that there exists a conserved gauge
current and, hence, a conserved electric charge, associated with fundamental
string sources, given by
\be
 e^B_2 = \int_{\Sigma_7}*\ e^{-\f}\ H\; ,
\ee
where $H=dB$ and $\Sigma_7$ is a compact 7-dimensional space. In our
example, this charge vanishes. However, there also exists a magnetic
charge, associated with solitonic $5$--brane sources,
\be
 g_6^B = \int_{\Sigma_3}H\; .
\ee
In our example there is a non-zero magnetic charge when the integral
is taken over the internal 3-dimensional subspace. Under the
non--compact scaling transformations in~\refs{form_scal},
\refs{g_scal} these conserved charges transform as 
\bea
 e^B_2&\rightarrow&\l^{-\frac{3r}{2}-\frac{s}{2}}\; e^B_2 \\
 g^B_6&\rightarrow&\l^{\frac{r}{2}+\frac{s}{2}}\; g_6^B\; ,\label{g_B}
\eea
respectively. Similarly, the invariance of the action under the Abelian
gauge transformation $A_\d\rightarrow A_\d +d\L_\d$ leads to two
conserved charges
\bea
 e^A_\d&=&\int_{\Sigma_{\tilde{\d}+1}}*\ e^{-\left(
          \frac{\d -4}{2}\right)\f}\ F_\d \\
 g_{\tilde{\d}}^A&=&\int_{\Sigma_{\d +1}}F_\d
\eea
associated with elementary $\d -1$ brane and solitonic $\tilde{\d}-1$
brane sources, respectively, where $F_\d = dA_\d$ and
$\tilde{\d}=8-\d$. Again, in our example the electric charge is zero,
but the magnetic charge is non-zero, when the integral is taken over
the internal 2-dimensional subspace. We find that
\bea
 e_\d^A&\rightarrow&\l^{-\frac{3r}{2}-\left(\frac{\d -4}{2}\right)
                   \frac{s}{2}}\; e_\d^A \\
 g_{\tilde{\d}}^A&\rightarrow&\l^{\frac{r}{2}+\left(\frac{(\d -4)}{2}\right)
                  \frac{s}{2}}\;  g_{\tilde{\d}}^A \label{g_A}
\eea
under the scaling transformation~\refs{form_scal}, \refs{g_scal}.

Let us first consider solutions for which all electric and magnetic
charges vanish. In this case, the effective four-dimensional 
theory governing these solutions
must exhibit the full scaling symmetry specified by $r$ and $s$. This
symmetry tells us that the dilaton potential must be flat, with any value
of the dilaton being an allowed vacuum. As we will see shortly, this is
indeed the case. Since the dilaton can take any value in this flat
potential, we conclude that this theory does not fix the dilaton vacuum.
Now consider solutions for which all electric and magnetic charges vanish
except for a single NS charge. To be specific, assume, as in the
example, that $g_6^B\neq 0$. All
solutions of the associated effective theory must preserve this charge.
Note from expression~\refs{g_B} that $g_6^B$ is, in general, not preserved
under scaling transformations. However, $g_6^B$ will be preserved under
the one--parameter subset of scaling transformations specified by
\be
 r=-s\; . \label{NS_cond}
\ee
It follows that the effective theory still must exhibit a scaling
symmetry, now specified by $s$ only. However, this reduced symmetry
implies that either the potential is flat, or it is non--flat with
no stable finite vacuum of the dilaton. In this case, as we will see
below, the vacuum degeneracy is lifted, but the dilaton runs off to
infinity. We conclude that this theory still cannot fix the dilaton
vacuum.

Now, however, consider the case that, in addition to the nonvanishing
NS solitonic charge $g_6^B$, there is also a non--vanishing RR solitonic
charge $g_{\tilde{\d}}^A$. We see from eq.~\refs{g_A} that this charge will
be scale invariant only if
\be
 r = -\frac{\d -4}{2}s\; . \label{RR_cond}
\ee
Since in a type IIA theory $\d\leq 3$, this expression is never compatible
with~\refs{NS_cond}. Therefore, scale invariance is completely
broken. It follows that the dilaton vacuum degeneracy must be 
lifted. This is a first, and necessary, step toward stabilizing the dilaton
vacuum. Again, it is not in itself sufficient because the vacuum may still run
off to infinity and never stabilize at a finite value. In this case, however,
it is possible that the dilaton potential has a finite, non--degenerate
minimum. We now show that the existence of two non-vanishing solitonic
forms, one NS and one RR, actually stabilizes the dilaton vacuum. Note that
the coefficients $a(\d_r )$ in action~\refs{action}, which control the
coupling of the dilaton to $\d_r$--forms, are given by $a=1$ for the
NS 2--form and $a(\d )=(\d -4)/2$ for a RR $\d$--form in type IIA.
Since $\d\leq 3$ for the RR form, it follows that $a(\d )<0$, opposite
in sign from the NS coefficient. Inserting the two solitons into
action~\refs{action} leads, in the reduced four-dimensional effective
theory, to a potential energy for the dilaton of the generic form
\be
 V_{\rm eff} = \frac{1}{2}\left(A^2e^{-\f}+B^2e^{\left(\frac{4-\d}{2}\right)
               \f}\right)\; , \label{Veff2}
\ee
where $A^2$ and $B^2$ are positive real numbers, related to the NS and
RR form magnetic charges. For theories with no
charges, $A=B=0$ and the dilaton potential
is flat, as we argued from scaling invariance. Theories with all
charges zero except $g_6^B$ have $A\neq 0$, 
$B=0$, thus admitting the first term in eq.~\refs{Veff2} only. The
potential is no longer flat, but the dilaton runs off to infinity. 
However, when $g_6^B$ and
$d_{\tilde{\d}}^A$ are non--vanishing, both $A$ and $B$ are non--zero, and 
the potential has a stable vacuum at
\be
 <\f > = \frac{2}{6-\d}\ln\left(\frac{A^2}{B^2}\frac{2}{4-\d}
         \right)\; .
\ee
Setting $\d =1$ yields the potential and the dilaton vacuum of the above
example, given in~\refs{V1_ex} and \refs{Vmin_ex}, respectively.
We conclude, that theories of this type exhibit a stable
dilaton vacuum. The reason for this stability is first, the complete
breaking of scale invariance by the topological charges, which must be
conserved, and second, the
fixing of the dilaton vacuum at a finite value due to the different sign
of the dilaton coupling to the NS and RR forms.

\section{Stabilizing moduli in M--theory}

Next, we would like to discuss the case where the dilaton
is viewed as a modulus. This is the appropriate point of view if one
considers M--theory where the dilaton arises as the compactification
radius of the eleventh dimension. Since the dilaton does not play a
special role from that perspective, there is no reason why one
should invoke a nonperturbative mechanism to stabilize the moduli but not
the dilaton. We should therefore ask the more ambitious question
whether the potential provided by the forms and curvature terms allows
for a stabilization of {\em all} moduli $\a_i$. Of course, there are
other moduli, corresponding to further deformations of the spherical
or toroidal subspaces and zero modes of the form fields, which we
have not included. Thus, in the examples that follow, we will
strictly be searching for solutions which stabilize a subset of the
moduli, including, hopefully, the dilaton.

The low-energy limit of M--theory is 11-dimensional supergravity,
which contains a single three-form potential. The action is of
our general form~\refs{action}, if we take $D=11$ and set the dilaton
and cosmological constant to zero. The full supergravity action
also includes a Chern-Simons term, describing the self-coupling 
of the three-form field. However, as stated above, for all the
configurations we will consider the contribution from this term is
zero, and so we can drop it from the action. That there might be
solutions with all moduli stabilized, is suggested by the original
seven-sphere compactifications of eleven-dimensional supergravity
discussed by Duff~\cite{7-sphere}. 
In these solutions spacetime is a product of a
four-dimensional anti-deSitter space and a seven-sphere of fixed
radius. The three-form potential is excited, so that the corresponding
four-form field strength spans the four-dimensional space. The
radius of the seven-sphere is directly related to the charge of the
form field. Considered as a compactification to four-dimensions, the
radius is a modulus field which appears to have been stabilized by the
presence of the three-form charge. 

We can see this stabilization directly by rewriting the solution in
the framework given at the beginning of this paper. By doing so we
will also show that it can sensibly be interpreted as the asymptotic
limit of a dynamical cosmological solution. That is, if the radius of
the seven sphere is a little away from its stabilized value, there is
a smooth solution where the external four-space continues to evolve
while the radius settles down into the minimum. First we note that
the space has been split into an external three-space and an internal
seven-space, so that $\bar\bd=(3,7)$ and we have a metric of the form
\be
  ds^2 = -\bar{N}^2(\t)d\t^2 + e^{2\bar\a_0}d\O_{K_0}^2
         + e^{2\bar\a_1}d\O_{K_1}^2 \; ,
\ee
where $\bar\a_0$ describes the curvature of the external space while
$\bar\a_1$ is the modulus describing the radius of the internal
space. Since the internal space is a seven-sphere, we
must have $K_1=1$, while the external space is anti-de Sitter, so must
have a negatively curved spatial subspace implying $K_0=-1$. These
curvatures contribute to the effective potential given in
eq.~\refs{U}, with $\bar\bq_r$ vectors given by $\bar\bq_{K_0}=(6,12)$ and
$\bar\bq_{K_1}=(4,14)$ respectively. The different signs of the
curvatures imply that the terms in the effective potential also differ
by a sign. For the seven sphere the coefficient of the exponential is
$u_r^2=-2$, while for the external three space $u_r^2=2$.  The form
field spans the external space and has a time-like component. Thus it
corresponds to an elementary Ansatz, and gives the vector
$\bar\bq=(6,0)$. Collecting all this together we find that the
effective potential~\refs{U} is given by
\be
  U = \frac{1}{2}u^2e^{6\bar\a_0} - 2e^{6\bar\a_0+12\bar\a_1} 
      + 2e^{4\bar\a_0+12\bar\a_1} \; ,
\ee
the three terms corresponding to the form field, the seven-sphere and
the curvature of the external space respectively. As discussed above,
we must make a Weyl rescaling in order to put the Einstein-Hilbert
action for the external, four-dimensional part of the metric in
canonical form, and so diagonalize the metric $G$ in the $\bar\bal$
space. The general transformation is given in eq.~\refs{trafo} and
here simply corresponds to introducing $\a_0=\bar\a_0+\frac{7}{2}\bar\a_1$
and $\a_1=\bar\a_1$. The potential can then be written as
\be
  U = e^{6\a_0}\left( \frac{1}{2}u^2e^{-21\a_1} - 2e^{-9\a_1} \right)
      + 2e^{4\a_0} \; . \label{duffU}
\ee
We note first that the last term, which comes from the curvature of
the external space, now no longer depends on the internal modulus
$\a_1$. This is as is expected since it is a property of the
external space alone. The other two terms provide a potential for
$\a_1$. The point here is that this potential has a minimum, which
fixes the radius of the internal sphere at
$\a_1=<\a_1>=\frac{1}{12}\ln(7u^2/12)$. 
There is a balance between the contribution to the potential
{}from the curvature energy of the internal seven-sphere, which
increases with radius, and the field energy due to the form field,
which decreases with radius. This balance is the origin of the
stabilization of the radius modulus. It is important that, in
calculating the dependence of the form-field energy on radius, we
recall that the charge of the solution cannot change dynamically. This
translates into the condition that the flux of the form field across
the seven-sphere, that is $\int_{S^7}*F$, which is proportional to
$u$, must remain fixed. 

To really show stability we must be a little more careful because there
is a dynamical prefactor $\exp(6\a_0)$ in the relevant terms in the
potential~\refs{duffU}. Following our previous discussion, we must
write out the equation of motion for $\a_1$ in comoving time,
defined by $N=\exp(7\a_1/2)\bar{N}=1$. From eqs.~\refs{V_def}
and~\refs{moduli} we find that the relevant potential, which is just the
effective potential in the reduced four-dimensional theory, is then 
\be
  V = \exp(-6\a_0) U = \left( \frac{1}{2}u^2e^{-21\a_1} - 2e^{-9\a_1} \right)
      + 2e^{-2\a_0} \; ,
\ee
and the prefactor disappears. Thus we can conclude that minimizing
the term in parentheses truly represents a stabilization of the radius of
the internal space. To complete the description of the solution, we
note that, at the minimum, the value of this term is negative and so
provides a negative cosmological constant. This is the reason why the
solution for $\a_0$ then gives a four-dimensional anti-de Sitter
space. 

Two further comments are worth making about this
solution. First, it is completely supersymmetric, preserving the full
$N=8$ supersymmetry in four dimensions. Secondly, it also represents
the infinite throat inside the membrane solution of eleven-dimensional
supergravity, as first discussed by Gibbons and
Townsend~\cite{throat}. In fact, as we will discuss below, there are a
number of other $p$-brane solutions with an infinite throat which lead
to cosmological solutions with stable moduli. 

\vspace{0.4cm}

While compactifying on a seven sphere provides an interesting example
of a pure supergravity solution with stable moduli it is not very
physical. It is more natural to look for solutions which have one
internal direction compactified on a circle. This radius can then be
related to the dilaton of string theory. 

Let us assume that the spacetime is split into an external $(3+1)$--space
with a scale factor $\a_0$ and an internal seven-space which is further
split into maximally-symmetric subspaces, one of which is a
circle. Any stabilization of the moduli will be controlled by the
potential $U$. Referring to eqs.~\refs{V_def} and~\refs{moduli}, we
recall that the potential that actually enters the canonical moduli
equations of motion is $V=\exp(-6\a_0)U$, the effective potential in
the reduced four-dimensional theory.  To be sure of a stable
solution, the part of $V$ which has a minimum for the moduli must be
independent of the external scale factor $\a_0$. This is equivalent to
the statement that, in the reduced effective four-dimensional theory,
the excited form field strengths appear either as 0-forms or as
4-forms, and so are not dynamical, but contribute only to the
effective potential. They correspond to the ``electric'' and
``magnetic'' configurations discussed in section two. One also notes that
the components of the vectors $\bq$ which control the exponentials in
the sum of terms which enter $U$, are always negative or zero. This
implies that it is impossible to stabilize all the moduli with a
such a potential unless at least one of the coefficients
$u_r^2$ is negative. The only way this is possible is to have some
positive curvature in the internal space. 

Having made these general observations let us consider a simple case
where we split the internal space into two three-spheres (providing
the necessary curvature) and a circle. In addition, we include two
solitonic orientations of the form field. One spans one three-sphere
and the circle, the other spans the other three sphere and the
circle. We also include an external space curvature, $K=0,\pm 1$,
in case the stable solution leads to a non-zero four-dimensional
cosmological constant. We will write $\a_1$, $\a_2$ and $\a_3$ for the
moduli of the two three-spheres and the circle respectively, and keep 
$\a_0$ for the external space. Using the expressions for the relevant
$\bq$ vectors~\refs{el_new}, \refs{sol_new}, \refs{curve_new} and
\refs{curvi_new}, we find that the potential $V$ is given by 
\be
  V = \frac{1}{2}u_1^2e^{-9\a_1-3\a_2-3\a_3}
      + \frac{1}{2}u_2^2e^{-3\a_1-9\a_2-3\a_3}
      - 2e^{-5\a_1-3\a_2-\a_3} - 2e^{-3\a_1-5\a_2-\a_3}
      - 2Ke^{-2\a_0} \; ,
\ee
where $u_1$ and $u_2$ are the charges of the two solitonic 
orientations of the form field. The last term represents the external
space curvature and does not effect the stabilization. To see if there
is a minimum of $V$, it is convenient to introduce new variables 
\be
  x = 2\a_1 + 2\a_2 + \a_3 \; , \qquad
  y = \a_1 - \a_2 \; , \qquad
  z = \a_3 \; .
\ee
The potential then reads
\be
  V = \frac{1}{2} e^{-3x} \left( u_1^2e^{-3y} + u_2^2e^{3y} \right)
      - 2e^{z-2x} \left( e^y + e^{-y} \right) - 2Ke^{-2\a_0} \; .
\ee
It is then clear the potential is not stabilized in the $z$
direction but rather goes to negative infinity as $z$ increase. For
fixed $z$ there is however a minimum in $x$ and $y$. In this sense the
potential ``stabilizes'' two of the moduli. 

This is, in fact, a generic result. Using our simple Ansatz with
maximally symmetric subspaces, it is not possible to find a solution
with one modulus describing a circle and all the moduli
stabilized. However, as in the case of the string theories discussed
previously, if one of the moduli gets stabilized by some other
mechanism, the presence of non-trivial form fields can then lead to
stabilization of all the other moduli. Two further points are worth
making. First, we only chose configurations which did not excite the
Chern-Simons term in the supergravity action. It is possible that
relaxing this condition provides the freedom necessary to stabilize
all the moduli. Secondly, in the most physical scenario,
corresponding to strong coupling limit of the heterotic string, one
dimension is compactified on an orbifold rather than a circle, and,
further, the presence of gauge fields living on the ten-dimensional
fixed points of the orbifold leads to sources for the form
field~\cite{witten}. Including either these effects would take us
outside the types of field configurations considered here. 

\vspace{0.4cm}

Even within our Ansatz of maximally-symmetric subspaces, many other
solutions with stable moduli exist, especially if one relaxes the
condition that the external space is four-dimensional. Some of these
solutions preserve some fraction of the supersymmetry. As an example,
consider a spacetime split into an external three-dimensional space
and an internal space which is the product of a three-sphere with a
four-torus and a circle. We excite a solitonic form across the sphere
and the circle and a fundamental form across the external space and the
circle. Let us assume that the external space is negatively curved. If we
write $\a_0$ for the external scale factor and $\a_1$, $\a_2$ and
$\a_3$ for the scale factors of the internal sphere, torus and circle
respectively, then, using eqns.~\refs{el_new}, \refs{sol_new},
\refs{curve_new} and \refs{curvi_new}, we find the effective
three-dimensional potential is given by
\be
  V = \frac{1}{2}u_1^2 e^{-12\a_1-8\a_2-4\a_3}
       + \frac{1}{2}u_2^2 e^{-12\a_1-16\a_2-2\a_3}
       - 2e^{-8\a_1-8\a_2-2\a_3} + 2e^{-2\a_0} \; .
\ee
Here $u_1$ is proportional to the charge of the solitonic form, while
$u_2$ is proportional to the charge of the fundamental form. It is
convenient to introduce two new variables
\be
  x = 4\a_2-\a_3 \; , \qquad z = 4\a_1+4\a_2+\a_3 \; ,
\ee
so that the effective potential can be rewritten as
\be
  V = \frac{1}{2} e^{-3z} \left( u_1^2 e^x + u_2^2 e^{-x} \right) 
       - 2e^{-2z} + 2e^{-2\a_0} \; .
\ee
Thus we find that the potential depends only on two of the three
moduli; that is to say, there is a flat direction. Moreover there is a
stable minimum at
\be
  x = \ln\left|u_2/u_1\right| \;, \qquad
  z = \ln\left|3u_1u_2/4\right| \;.
\ee
Thus two of the three moduli are stabilized while the third corresponds
to a flat direction and so can take on any constant value, implying
that we have a consistent cosmological solution with fixed moduli.
The value of $V$ at this minimum is negative so that the external space
is in a de Sitter phase. 

Furthermore, this solution corresponds to the infinite throat
of the intersecting membrane-fivebrane solution of 
M--theory~\cite{multibrane}, in the degenerate limit
where the brane charges are equal (so we take $u_1=u_2$). As such it preserves
one-quarter of the supersymmetry. Other solutions can similarly be
identified with,  for instance, the infinite throats of the single
fivebrane solution and the degenerate triple fivebrane solution,
preserving all and one-eighth of the supersymmetry
respectively. Likewise, there are cosmological solutions corresponding
to the throats of $p$-brane solutions in type II and heterotic
solutions. Again, they describe the stabilization of several moduli,
though, in general the dilaton is not fixed in these
solutions. Similar behavior, of an effective potential with enhanced
supersymmetry at points where the moduli become fixed, has been
observed in work on black holes~\cite{fk,kr,cal,kl}.

\section{Conclusion}

In this paper, we have shown that non-trivial form fields of type II and
M--theory provide an effective potential for the dilaton and moduli, which 
can fix these fields to a finite, stable 
minimum during their cosmological evolution. The value of the fields
at the minimum is controlled by the strength of the form field
charges. The structure of this potential
is such that after a
short period of oscillations around the minimum, which are damped by the
expansion of the universe and a possible decay of the coherent modes,
the moduli and dilaton settle down to what should be interpreted as
the vacuum of low-energy particle physics. Furthermore, this process
is consistent with an ongoing expansion of the observable universe.

More specifically, we have addressed cosmological dynamics in type II
theories. The dilaton can be fixed by
turning on solitonic NS and RR forms in the internal space, once the
geometrical moduli are stabilized by an additional nonperturbative potential.
A physical understanding of this can be obtained by analyzing the scaling
symmetries of the theory. We argued that these symmetries, which normally
prevent a dilaton stabilization, are broken by the conserved form
field charges. 

In M--theory, we asked the more ambitious question of whether the
moduli can be consistently 
fixed without invoking additional nonperturbative effects. It turned out
that this is indeed possible in simple examples by turning on solitonic
forms in the internal space (or an elementary form which covers the full
external space) and by using positively curved internal spaces. Moreover,
these examples show that part of the supersymmetry, which we generically
expect to be completely broken during the early period of the moduli
evolution, can be restored once the moduli have settled down to their
vacuum. Therefore our mechanism can be consistent with the idea of low
energy supersymmetry.

Some properties of this cosmological scenario are reminiscent of phenomena
observed in the context of string black holes~\cite{fk,kr,cal,kl}.
There it has been noted that certain scalar fields are attracted to fixed
points once the radial coordinate approaches the black hole
horizon. Moreover, at these fixed points supersymmetry is restored. These
analogies between the time evolution of cosmological models and the radial
dependence of black holes are not surprising given the fact that a
subclass of the types of cosmological solutions we consider corresponds to
the interior solutions of black p--branes where the radius coordinate
becomes timelike~\cite{paper,be_fo,rudi,LW}.
However, it should be stressed that the mechanism discussed in this paper
is not restricted to cosmological models related to black holes, but
applies to a much wider class.

\vspace{0.4cm}

{\bf Acknowledgments} A.~L.~is supported by a fellowship from
Deutsche Forschungsgemeinschaft (DFG). A.~L.~and B.~A.~O.~are
supported in part by DOE under contract
No. DE-AC02-76-ER-03071. D.~W.~is supported in part by DOE under
contract No. DE-FG02-91ER40671.
\end{document}